\documentstyle[aps,preprint]{revtex}
\begin{document}
\draft
\title{Microwave-induced
control of free electron laser radiation.}
\author{A. J. Blasco$^1$, L. Plaja$^1$ , L. Roso$^1$, F. H. M.
Faisal$^2$}
\address{(1) Departamento de F\'\i sica Aplicada, Universidad de
Salamanca,\\ E-37008 Salamanca,
Spain}
\address{(2) Fakult\"at f\"ur Physik, Universit\"at Bielefeld, Bielefeld,
D-33501
Germany}
\date{\today}
\maketitle
\begin{abstract}
The dynamical response of a relativistic bunch of
electrons injected in a planar magnetic
undulator and interacting with a counterpropagating
electromagnetic
wave is studied. We
demonstrate a resonance condition for which the free electron laser
(FEL)
dynamics is strongly influenced
by the presence of the external field. It opens up the possibility of control of
short wavelength FEL emission characteristics by changing
the parameters of the microwave field without
requiring change in the
undulator's geometry or configuration.
Numerical examples, assuming realistic parameter
values analogous to 
those of the TTF-FEL, currently under development at DESY,
are given for possible control
of the amplitude or the
polarization of the emitted radiation.

\end{abstract}

\narrowtext

\section{Introduction.}

Since their first experimental realization \cite{elias1},
free
electron lasers (FEL) have been 
one of the most promising sources of coherent electromagnetic
radiation
\cite{dattoli,fel}. The physics of FEL emission is radically
different from that of
any other laser sources. In
particular, the tunability
over a broad range of frequencies, and the brightness of its output are difficult
to
achieve in other lasing schemes. On the other hand, its
polarization, pulse shape, etc. are strongly
connected with the
geometry of the undulator and hence are inconvenient to modify. 
At the same time, modifications in the typical undulator's physical structure
may induce new features, e.g., FELs with two
magnetic wigglers of
different spatial frequencies may increase the radiation at higher harmonics
\cite{schmitt},
suppress the side bands
\cite{iracane}, and may allow the radiation spectrum \cite{kong} to
be
controlled. However, a systematic experimental exploration of these
possibilities is very ackward, if
not
precluded, due to the difficulty of engineering and constructing the
modified undulators 
for every such
experiment. It is therefore worthwhile to explore
theoretically the possibility of achieving control of the
amplitude
and polarization of the emitted radiation, specially at
very short wavelengths, without having
to
alter the undulator geometry. 

The basic dynamics of the interaction of free electrons interacting
with
electromagnetic 
waves has been studied in many circumstances in the past. They range from
pioneering studies
of the radiation of a single electron driven by an
electromagnetic wave
\cite{brown,sarachik}, interaction of
relativistic electrons under
general initial conditions with such
radiations
\cite{salamin1}, charged particle
acceleration by
simultaneous interaction 
with an electromagnetic wave and a static electric
field
\cite{hussein,salamin2}, etc. 
Another important application of FEL principle is the
particle
acceleration by the inverse mechanism. 
Particle
acceleration 
by inverse free-electron-laser
principle has been demonstrated both theoretically
\cite{courant} and experimentally
\cite{wernick,gallardo}.
In contrast, much less seems to be known for the
complementary geometry, in which the
electron bunch interacts with a
{\em counterpropagating} electromagnetic wave, perhaps because of the absence
of
acceleration schemes for this case. 
In this paper we explore the possibility of modifying the electron
dynamics in the amplification
stage of FEL, by means of the interaction with a counterpropagating
microwave
field. It will be shown that, under certain conditions, the counterpropagating wave can
influence strongly the
dynamics of the electrons inside the
undulator. Thus, by a
careful choice of the wave parameters control of the
dynamics can be
achieved that could lead to desired FEL radiation properties 
without requiring geometrical
changes of the undulators. 

\section{Electron dynamics and Phase matching
condition}
\label{sec:electron}

Let us consider a modified FEL configuration as depicted in Fig.
\ref{fig:geometry}: a
free electron is injected axially into a linearly polarized magnetic undulator, where
an
electromagnetic wave propagates also axially, in opposite direction, inside a
waveguide. The evolution of
the electron motion, in the combined steady
magnetic field of the undulator and the electromagnetic wave, is
governed by
the Newton-Lorentz  equation
\begin{equation}
\label{eq:lorentz} {{d ~} \over {d t}} \vec{p}=q
\left[ \vec{E}_0 + {1 \over c}
\vec{v} \times
\left(\vec{B}_u +\vec{B}_0
\right) \right]
\end{equation} where
$\vec{B}_u= f(x) B_u  \sin(k_u x) \vec{e}_z$ is the undulator's
magnetic field. The explicit form of the
counterpropagating field depends on the waveguide geometry, as
well as on the choice of a particular transversal
mode. Due to the small transversal dimensions of the
electron bunch used in FEL (of about some tens of microns),
only the field at the central axis of the
waveguide is relevant. The following discussion, therefore, may be
applied to any waveguide mode of any
geometry, provided it has a non vanishing linearly polarized field
along the central axis, which can be regarded
as constant over the whole bunch's section. The choice of particular
waveguide parameters will
influence the quantitative values where the interference condition considered below is
attained.
For concreteness, let us assume a $TE_{n0}$ \cite{ramo} mode propagating in a rectangular
waveguide.
The explicit forms for the electric and magnetic field now read as
follows,
\begin{equation}
\label{eq:Ewaveguide}
\vec{E}_0=- E_0 g(k_w x+\omega_0 t) {k_0 \over k_c} \cos(k_c
z) 
 \sin(k_w x+\omega_0 t + \phi_0) \vec{e}_y
\end{equation}

\begin{eqnarray}
\label{eq:Bwaveguide}
\vec{B}_0 & = & - E_0 g(k_w x+\omega_0 t)
\sin(k_c z)
\cos(k_w x+\omega_0 t+ \phi_0) \vec{e}_x + \nonumber
 \\
&&E_0 g(k_w x+\omega_0 t) {k_w \over
k_c} \cos(k_c z) 
\sin(k_w x+\omega_0 t + \phi_0)
\vec{e}_z
\end{eqnarray}
$k_w= \sqrt{k_0^2- k_c^2}$ being the wave number of the travelling wave, $k_0=
{\omega_0 \over c}$, and
$k_c={{n \pi} \over a}$ the cutoff wave number of the waveguide ($a$ being
the width
of the waveguide). $f(x)$ and
$ g(k_w x+\omega_0 t)$ are considered slowly varying envelopes.

Without loss of
generality (by shifting the time coordinate)  we may assume that the
electron is initially at $x=0$, moving
along the $x$-axis with velocity
$v_0$. Before solving the equation of motion numerically, we may gain a
qualitative
insight into the problem by first considering the dynamics in a new reference frame in
which the
electron is initially at rest. In the new frame, the  undulator's magnetic
field becomes a counterpropagating
time-varying electromagnetic field with
\begin{eqnarray}
\vec{E}_u' & = & - \gamma \beta f(k_u' x'+\omega_u'
t') B_u  \sin(k_u' x'+\omega_u'
t') \vec{e}_y \\
\vec{B}_u' & = & \gamma f(k_u' x'+\omega_u' t') B_u \sin(k_u'
x'+\omega_u' t')
\vec{e}_z
\end{eqnarray} 
where $k_u'=k_u \gamma$, $\omega_u'=k_u' v_0$  and $\gamma
=
1/\sqrt{1-\beta^2}$ being the Lorentz factor ($\beta = v_{0}/c$). Note that
this electromagnetic field has
two peculiarities: one, the magnetic and the electric field
amplitudes do not coincide in their strengths, the
electric field being smaller, and
two, it propagates with a velocity $v_0 = \omega_{u}^{'}/k_{u}^{'} < c$. A
more
fruitful way is to reinterpret this field as an electromagnetic wave propagating in
vacuum with a space
dependent phase
\begin{eqnarray}
\label{eq:eup}
\vec{E}_u' & = & - \gamma \beta B_u f(\kappa_u' x'+\omega_u'
t'+
\phi_u(x')) \sin(\kappa_u' x'+\omega_u' t'+ \phi_u(x'))
\vec{e}_y
\\
\label{eq:bup}
\vec{B}_u' & = &
\gamma B_u f(\kappa_u' x'+\omega_u' t'+ \phi_u(x'))
\sin(\kappa_u' x'+\omega_u'
t'+
\phi_u(x'))
\vec{e}_z
\end{eqnarray}
\begin{equation}
\label{eq:fase}
 k_u' x'+\omega_u' t'=\kappa_u'
x'+\omega_u' t'+ \phi_u(x')
\end{equation} 
with $\kappa_u'=\omega_u'/c$ and $\phi_u(x')=\omega_u' \left( {1
\over
v_0} - {1 \over c} \right) x'$.

On the other hand, the counterpropagating electromagnetic wave in the
new
reference frame becomes
\begin{eqnarray}
\label{eq:e0p}
\vec{E}_0' & = & - \gamma E_0 {{k_0 +k_w
\beta}\over k_c} cos(k_c z') g(k_w'
x'+\omega_0' t')
\sin(k_w' x'+\omega_0'
t'+
\phi_0)
\vec{e}_y
\\
\label{eq:b0p}
\vec{B}_0' & = & - E_0 sin(k_c z') g(k_w' x'+\omega_0'
t') cos(k_w'
x'+\omega_0' t'+\phi_0) \vec{e}_x+ \nonumber
\\ &&\gamma E_0 {{k_0 \beta +k_w} \over k_c} cos(k_c z') g(k_w'
x'+\omega_0'
t')
\sin(k_w' x'+\omega_0' t'+
\phi_0)
\vec{e}_z 
\end{eqnarray} with $\omega_0'= \gamma (
\omega_0 +k_w \beta c)$ and
$k_w'=\gamma (k_w + k_0 \beta)$.
Note that in the strong relativistic case (large
$\gamma$), the effective field observed by the
electron can be considered as a $TEM$ wave. In addition, 
the
phase velocity of this field approaches 
$c$. These two facts, together with the condition $k_c z'\simeq 0$,
which is ensured by the reduced
dimensions of the electron bunch, permit us to ascribe the effective field
acting on the electron
in its rest frame to a plane wave.

By inspection of Eqs. (\ref{eq:eup}-\ref{eq:b0p}),
one sees that it is
possible to derive a phase matching condition in which both fields can be
seen to have the
same frequency in the moving frame,
\begin{equation}
\label{eq:phase_matching}
\left.
\matrix{
{k_w'=\kappa'_u} \cr {\omega_0'=\omega_u'} \cr}
\right \} \rightarrow {\omega_0 \over c}= { {k_u^2 +k_c^2}\over
{ 2 k_u}} 
\end{equation}
provided that the waveguide has a transverse dimension greater than half
the
undulator's wavelength, $a>{\lambda_u \over 2}$.
 Note that condition (\ref{eq:phase_matching}) has been
calculated
for the case of strongly relativistic electrons, $\beta
\approx 1$. 
Although this result is
derived for a rectangular waveguide, it is
worth to stress that this is independent
of the particular geometry
(which is described by the appropiate form
of $k_c$). Note also that the dependence of the frequency of the
electromagnetic wave on $k_c$
permits to attain the same phase matching condition for a variety of
electromagnetic waves, only by
modifying the waveguide geometry. 

 As given by Eq. (\ref{eq:phase_matching}),
the phase matching condition is
defined only for the temporal oscillation. Since the undulator field has
a
spatial phase dependence, the corresponding wavenumber matching will hold
only over a certain coherence
length such that
$\phi_u(\ell_{coh}')=\pi$,
\begin{equation}
\label{eq:coherence_length}
\ell_{coh}'={
{\lambda_u/2}\over {\gamma (1-\beta)}}
\end{equation}
Before proceeding further, we may point out  that the
nature of the
electron motion can strongly depend on this coherence length and show an
interesting disordered
behavior when the coherence length becomes comparable to the
undulator's wavelength. However, here we are
concerned with the condition in which the
coherence length is greater than the total undulator length. This  
condition is
easily fulfilled by very high energy electrons.  In this situation the motion of the
electron
remains regular.

\section{Numerical Integration}
\label{sec:method}

As indicated  above our 
objective is
to study the possibility of modifying the
FEL emission characteristics induced by
the counterpropagating
microwave field, 
in a configuration
similar to that being developed in DESY \cite{saldin,treusch}. The initial
conditions, hence, consist
of a relativistic electron-{\it bunch} entering the undulator 
in the presence of  a
very weak
{\em seed} of FEL radiation field, which is assumed to be generated from vacuum noise in a
first
stage of the FEL laser. Since the bunch injection energy is high, the dynamics
encloses two very
different space-time scales, namely, that of the undulator's field and that
corresponding to the output
radiation, which differ typically by a factor $\gamma^2$. This disparity
becomes a limiting difficulty for the
numerical integration of the evolution equations, which is
usually overcome by using the appropriate slowly
varying envelope
approximations, along with the
projections on the field cavity
modes
\cite{dattoli2,goto,sprangle}. In this work we have chosen an alternative procedure
\cite{elias2} which
computes the radiated field from the superposition of the {\em
Li\'enard-Wiechert} fields \cite{jackson} emitted
from every
pseudoparticle (see below) of the bunch. Furthermore, we have
preferred to integrate the equations
in the initial rest frame of the
bunch. This allows us to avoid the problem associated
to the disparity of
scales since in this frame the undulator and radiated field have similar
frequencies. Moreover, in the chosen
frame, it
becomes readily evident that the bunch density
is decreased by a factor $\gamma \gg 1$, allowing us
to neglect self-fields.
The large number of electrons per bunch (in our case $\simeq 10^9$) in
the realistic
situation
forces us to define {\em
pseudo}particles each of which include a few thousands of electrons
that
are assumed to move 
together. Note that this is the same conceptual philosophy as employed in the
successful
particle-in-cell (PIC) codes for the simulation of
plasma dynamics \cite{dawson,birdsall}. 
The modulations of
the charge density in the system can be modeled either by
considering the
spatially variable distribution of
equally charged pseudoparticles, or by a
spatially uniformly distributed
set of variably charged
pseudoparticles. 
For convenience, we have chosen the latter approach in the present
investigation. To simulate
the velocity and acceleration of the
pseudoparticles, we have used a relativistic Boris algorithm
\cite{birdsall}
 and to calculate the resulting emitted field
of the electrons in the forward direction, we
have used the well-known
formula of the far field radiation field amplitude of an accelerated
charged
particle
\cite{jackson}:
\begin{equation}
\label{eq:E_radiated}
\vec{E}_{rad}(t)= \left.{q \over c}
{{\dot{\beta}_y(t')- \beta_x(t')
\dot{\beta}_y(t')-\beta_y(t) \dot{\beta}_x(t)}
\over
{(R-x(t'))(1-\beta_x(t'))^3}} \right|_{t'=t-(R-x(t'))/c} \vec{e}_y
\end{equation} 
where $R$ is assumed to be
large enough.
Once the integration is performed, we Lorentz-transform the computed quantities to the
laboratory
reference system.

\section{Coherent Control of FEL Radiation}
\label{sec:results}

In this
section we will theoretically demonstrate the possibility of controlled FEL
radiation through the external
electromagnetic wave. The key idea is to consider a
counterpropagating wave resonant with the undulator field,
in the sense discussed in section
\ref{sec:electron}. The frequency of the wave depends, therefore, on the
spatial
periodicity of the undulator's magnetic field and on the particular geometry of the waveguide. In
our
case, we take the
$2.73 cm$ undulator's wavelength of TTF-FEL at DESY \cite{treusch}, and a $TE_{10}$ mode
of a
rectangular waveguide of a size of $1.5 cm$ , which is similar to the size of the beam pipe of the
FEL at
DESY. Equation (\ref{eq:phase_matching}) defines the resonant condition for a
counterpropagating electromagnetic
wave in the microwave region with $\lambda=2.99 cm$ when
propagating in free space.

In addition to a resonant
frequency, the microwave control of the FEL amplification is 
more effective for the case in which the amplitude
of this field in the bunch's rest frame equals
the amplitude of the electromagnetic wave associated to the
undulator's field. At
present, microwave fields in the G$Hz$ range are available with powers up to $100 MW$
\cite{hanjo}.
Although this is already close to the value needed to control optimally the radiation of
the
TTF-FEL at DESY, we prefer to be conservative and to consider in this paper a tapered undulator
to reduce
the undulator's magnetic field to 25 $mT$. With this values it should be possible to
demonstrate the
microwave-control experimentally with current technology. On the other hand, the
state of the art of the
microwave generation by the FEL concept allows to foresee the availability
of brighter sources in the near
future \cite{wang}.

Unless stated otherwise explicitly, the results are based on
calculations for
an
electron bunch (of 300 MeV) injected into a $4.5 m$
magnetic undulator, whose characteristics have been
commented upon
above. Our numerical tests show a posteriori that
the assumption of an initially cold bunch is
acceptable. The bunch is described by a
spatial $sin^2$ distribution of $20000$ particles, $250 \mu m$
long.
Small changes of this number and/or the choice of bunch shape is found not to affect the
conclusions drawn from
the simulations.

In the following, we consider two cases of microwave control of
free electron laser
emission. 
First, we will analyze the possibility of suppressing the FEL
output by microwave interaction,
opening ways to control the pulse
of the FEL 
radiation by modulation of the
microwave amplitude. 
Second,
control of the polarization angle of the FEL radiation
by  changing the microwave polarization.
These
possibilities are particularly interesting in view of the lack of
convenient optical elements at very short
wavelengths to manipulate
these characteristics of FEL radiation once they are extracted from
the
source.

\subsection{Coherent suppression of radiation}

Let the counterpropagating microwave field be
linearly
polarized, with the polarization vector perpendicular to the direction of the undulator's
magnetic
field. The undulator field and the microwave field may be, then, made to
interfere destructively when
they have their phases properly matched
at the a critical value of the
amplitude of the microwave
field,
\begin{equation}
\label{eq:E_critical} E_{crit} = {k_c \over { k_0 \beta +k_w}} B_u
\end{equation}

Note that (like the phase matching condition, Eq.
(\ref{eq:phase_matching}))  the critical field becomes
almost independent of the
energy of the electron in the highly relativistic case, $\beta \approx
1$. Thus, a
nearly complete
destructive interference can occur for the  sum magnetic field, or
$B_{T}'=B_u'+B_0' = 0 $, for
all time, in the moving frame. In
contrast, because of the asymmetry
between the magnetic and electric field
amplitudes of the (Lorentz-transformed)
undulator field, the total electric field in the moving frame, on the
other hand,
does not vanish exactly. A small residual electric field
$E_{res}'=\gamma (1-\beta) B_u$ remains,
which, however, diminishes greatly as the
electron energy increases. Thus for $\beta \approx 1$,
$E_{res}'\simeq
B_u/\gamma^2$, the interference condition for the total electric field becomes
almost exactly
fulfilled.

Fig. \ref{fig:energy} shows the resulting suppression of FEL
radiation calculated for 
different
initial bunch
energies. Note that, as expected, the amplification
gain is dramatically reduced as the bunch
energy increases. 
This is because the residual electric field 
, $E_{res}'$
vanishes with increasing energy
of the bunch. 
Note that, for the higher energies, the microwave field reduces the gain
by nearly 3 {\it
orders} of magnitude. 

For the destructive interference mechanism to be effective in practice,
it is required
a constant $\pi$ phase-difference between the undulator and the
microwave fields, as seen in the electron
bunch's reference system. 
Note that
ensuring an initial
$\pi$ dephase requires a certain control of the
bunch's conditions
before injection, since the bunch
must enter the undulator when the microwave phase is
opposite to the
undulator's. The required constancy is ensured 
by (\ref{eq:coherence_length}) 
since the
coherence length 
is greater than the undulator's dimension
for the range of
bunch energies assumed here (e.g., TTF-FEL). 

To analyze the sensitivity of the coherent suppression effect

against fluctuations in the initial field
dephasing, we have performed a series of calculations in which the
bunch's initial position against
the undulator vertex is changed. 
The shift in the initial position is
directly related to the time
delay of the bunch to reach the undulator and, therefore, to the
initial dephase
between the
undulator and microwave fields (in the rest frame of the bunch). 
The results of calculations with
a 
300$MeV$, 250$\mu m$  bunch are
presented in Fig. \ref{fig:mili} which shows the amplification
vs.
fluctuations in the bunch position. 
It can be seen that the gain suppression effect is robust against
fluctuations less than $1
mm$ that is well above the usual experimental uncertainty.

\subsection{Control of
polarization of FEL radiation}

Polarization control is of particular interest for very short 
wave FEL
radiation. This
can be achieved in the same configuration by
rotating the polarization of the microwave by a
certain angle from the
undulator's plane of polarization. From the theoretical analysis
above we expected that,
in
general, the emitted radiation will be elliptically polarized. The
ellipticity would depend on the
initial
dephase of the fields as well as their relative amplitudes.
We show in  Fig.
\ref{fig:ellipticity}a 
the
calculated change of ellipticity of FEL
output vs. the relative
angle between the polarization plane of the
undulator  
and that of the 
microwave (chosen to be linearly polarized). 
The amplitudes of the undulator and
microwave fields in the rest
frame of the bunch are chosen to be comparable 
while the initial dephase is set
to $\pi/2$. 
The resulting ellipticity of the emitted radiation 
is found (Fig. \ref{fig:ellipticity}a) 
to
change from the linear to the circular
polarization. The results of the various cases presented in this
figure
are summarized in Fig.
\ref{fig:ellipticity}b in terms of the tilt-angle of the major axis of
the
polarization-ellipse of the emitted radiation. 

\section{Conclusion.}

Coherent modifications of FEL
radiation induced by a {\it counterpropagating} 
electromagnetic wave interacting with
an electron bunch in a
magnetic undulator are studied. 
A phase-matching condition between the undulator field and an
external
microwave field in the rest frame of a relativistic electron bunch
is derived. This condition is found
to only depend on the geometry of the
problem.
 It is found that possible control of both the amplitude and
the
polarization of the FEL radiation (including very short wavelengths)
could be achieved without having to
alter the undulator's geometry,
by simply varying the incident microwave field. 
Results of concrete numerical
simulations assuming realistic FEL
parameters (corresponding to that of TTF-FEL, currently under
development at
DESY) are given, and their robustness  
against small fluctuations in initial conditions is
illustrated.

\pagebreak

\acknowledgments

We thank Luis Elias for useful discussions. L. P. wishes to
acknowledge with thanks support from
the Spanish Ministerio de Educaci\'on y Cultura (under grant EX98-35084508)
Partial support from
the Spanish Direcci\'on General de Ense\~nanza Superior e Investigaci\'on Cient{\'\i}fica 
(grant
PB98-0268), from the Consejer{\'\i}a de Educaci\'on y Cultura of the Junta de Castilla y Le\'on
(Fondo
Social Europeo), (under grant SA044/01) and from DFG,
Bonn, under SPP: Wechselwirkung intensiver
Laserfelder mit Materie, FA
160/18-2, are thankfully acknowledged.

\begin{figure}
\caption{Schematic diagram
of the modified FEL amplifier configuration used throughout this paper.
An electron-bunch is injected into the
linearly polarized
magnetic undulator in the presence of a
counterpropagating electromagnetic
wave.}
\label{fig:geometry}
\end{figure}

\begin{figure}
\caption{Dependence of the FEL amplification factor
versus the
initial bunch energy,
in the presence of the
microwave field (dashed line) and in its absence
(solid line).}
\label{fig:energy}
\end{figure}

\begin{figure}
\caption{Dependence of the amplification
factor versus the 
initial coordinate of the bunch; $x=0$ corresponds to the 
initial position for which the
undulator and the microwave fields in the rest frame of the bunch have
opposite phases.}
\label{fig:mili}
\end{figure}

\begin{figure}
\caption{ (a) Ellipticity of the amplified FEL
radiation versus the
relative angle between the planes of polarization of the
microwave and the undulator
fields. (b) Tilt-angle  of the major axis of
the polarization ellipse of the FEL radiation with respect to the
polarization plane
of the undulator field, for
the same cases as considered in
(a).}
\label{fig:ellipticity}

\end{figure}


\begin{references}

\bibitem{elias1} L. R. Elias et al.,
Phys. Rev. Lett., {\bf 36}, 717 (1976).

\bibitem{dattoli} For a review of the fundamentals of Free Electron
Lasers, see, e.g.,
G. Dattoli, L. Giannessi, A. Renieri and A. Torre, Prog. in Opt. XXXI, 321
(1993).

\bibitem{fel} A collection of state-of-the-art papers on
Free Electron Lasers, in J. Feldhaus,
H.
Weise, {\em Free Electron Lasers 1999} (Elsevier Science, Amsterdam,
2000).

\bibitem{schmitt} M.J.
Schmitt, C.J.Elliott, IEEE J. Quantum
Electron. {\bf 23}, 1552 (1987).

\bibitem{iracane} D. Iracane, P. Bamas
Phys. Rev. Lett {\bf 67}, 3086
(1991).

\bibitem{kong} M.G. Kong, X. Zhong, A. Vourdas, Nucl. Inst. and Meth.
A {\bf 445},7 (2000).

\bibitem{brown} L.S. Brown and T.W.B. Kibble, Phys. Rev A {\bf 133}, 705
(1965).

\bibitem{sarachik} E.S. Sarachik and G.T. Schappert, Phys. Rev. D {\bf 1},2738 (1970), and
references
therein.

\bibitem{salamin1} Y.I. Salamin, F.H.M. Faisal, Phys. Rev. A {\bf 54},
4383
(1996).

\bibitem{hussein} M. S. Hussein, M. P. Pato, and A. K. Kerman, Phys. Rev. A {\bf 46},3562
(1992).

\bibitem{salamin2} Y.I. Salamin, F.H.M. Faisal, Phys. Rev. A {\bf 58}, 3221
(1998).

\bibitem{courant} E.D. Courant, C. Pellegrini and W. Zakowicz, Phys. rev. A {\bf 32}, 2813
(1985).

\bibitem{wernick} I. Wernick and T.C. Marshall, Phys. Rev. A {\bf 46}, 3566 (1992).


\bibitem{gallardo} A. van Steenbergen, J. Gallardo, J. Sandweiss and
J.-M. Fang, Phys. Rev. Lett., {\bf 13},
2690 (1996).

\bibitem{ramo} S. Ramo, J. R. Whinnery and T. Van Duzer, {\em Fields and
waves in communication
electronics} (Wiley \& Sons, New York, 1994)

\bibitem{saldin} E. L. Saldin, E. A. Schneidmiller, M. V. Yurkov,
Nuclear
Instruments and Methods in Physics Research A {\bf 445}, 178 (2000).

\bibitem{treusch} R. Treusch,
{\em Photon Beam Properties of the VUV-FEL at DESY},
communication in VUV FEL User Workshop (DESY, March 1999).
Also in
http://www.hasylab.desy.de/.

\bibitem{dattoli2} G. Dattoli et al. J. Appl. Phys. {\bf 80}, 6589
(1996).

\bibitem{goto} M. Goto et al. Nuclear Instruments and Methods in Physics Research A {\bf 445},
45
(2000).

\bibitem{sprangle} P. Sprangle, A. Ting, C. M. Tang, Phys. Rev. A {\bf
36}, 2773
(1987).

\bibitem{elias2} L. R. Elias, I. Kimel, Nuclear Instruments and Methods in Physics Research A
{\bf

393}, 100 (1997).

\bibitem{jackson} J. D. Jackson, {\em Classical Electrodynamics} (Wiley \& Sons,
New
York, 1998).

\bibitem{dawson} J. M. Dawson, Rev. Mod. Phys. {\bf 55}, 403 (1983).

\bibitem{birdsall}
C. K. Birdsall and A. B. Langdon, {\it Plasma Physics via
Computer Simulation}, Plasma Physics Series (IOP
Publishing, Bristol, 1991).

\bibitem{hanjo} H. Hanjo, Y. Nakagawa, J. Appl. Phys. {\bf 70}, 1004
(1991).

\bibitem{wang} Wang Pingsham, Lei Fangyan, Huang Hua, Gan Yanqing, Wang
Wendou, Gu Binglin, Physical
Review Letters {\bf 80}, 4594 (1998).

\end{references}
\end{document}